\documentclass[twocolumn,showpacs,amsmath,amssymb,aps,prd,floatfix,longbibliography]{revtex4-2}  
\usepackage{amssymb,amsmath}
\usepackage{color}
\usepackage{graphicx}
\usepackage{dcolumn}
\usepackage{bm}
\usepackage{upgreek}
\usepackage{marvosym}
\usepackage{latexsym,epsfig} 
\usepackage{bm}
\usepackage{mathtools}
\usepackage[T1]{fontenc}
\usepackage{hyperref}
\usepackage[nodisplayskipstretch]{setspace}

\DeclareUnicodeCharacter{2212}{-}

\begin{document}
\title{Giant magnetic
and optical anisotropy in cerium-substituted M-type strontium
hexaferrite driven by 4$f$ electrons} 

\author{Churna Bhandari$^1$}
\email{cbb@ameslab.gov (corresponding author)}
\author{Durga Paudyal$^{1,2}$}
\affiliation{$^1$The Critical Material Institute, Ames National Laboratory, Iowa State University, Ames, IA 50011, USA
\\$^2$Department of Electrical and Computer Engineering, Iowa State University, Ames, Iowa 50011, USA}

\begin{abstract}
By performing density functional calculations, we find a giant magnetocrystalline anisotropy (MCA) constant in 
abundant element cerium (Ce) substituted M-type hexaferrite, in the energetically favorable strontium site, assisted by a quantum confined electron transfer from Ce to specific iron (2a) site. Remarkably, the calculated electronic structure shows that the electron transfer leads to the formation of Ce$^{3+}$ and Fe$^{2+}$ at the $2a$ site producing an occupied Ce($4f^1$) state below the Fermi level that adds a significant contribution to MCA and magnetic moment. A half Ce-substitution forms a metallic state, while a full substitution retains the semiconducting state of the strontium-hexaferrite (host). 
In the latter, the band gap is reduced due to the formation of charge transferred states in the gap region of the host. The optical absorption coefficient shows an enhanced anisotropy between light polarization in parallel and perpendicular directions. Calculated formation energies, including the analysis of probable competing phases, and elastic constants confirm that both compositions are chemically and mechanically stable. 
With successful synthesis, the Ce-hexaferrite can
be a new high-performing critical-element-free permanent magnet material adapted for use in devices such as automotive traction drive motors.

\end{abstract}
\maketitle

\section{Introduction}
The increasing demand for high-performance permanent magnets for higher energy efficiency miniaturized devices with the limited global supply of critical rare-earth elements has urged the magnetism community to discover an alternative to critical rare-earth-based magnets. Although the hexaferrites are useful critical rare-element-free permanent magnets, the low {\color{black} value of maximum energy product} (BH$_{\text{max}})$\cite{BH} limits their usage for high-performance magnetic devices. Continuous theoretical and experimental efforts have been made to improve their intrinsic magnetic properties. In particular, an electron doping via $5d$-element La substitution is a promising approach in uplifting the magnetocrystalline anisotropy (MCA) constant ($K_1$) of ${\rm SrFe}_{12}{\rm O}_{19}$\cite{LotgeringJPCS74, SeifertJMMM09}, albeit it reduces the net magnetic moment. The magnetic rare-earth elements are desirable over La because a large spin-orbit coupling (SOC) in $4f$-states changes the MCA drastically, as seen in other rare-earth-based permanent magnets\cite{Rehman019, Chang012, MoslehJJM016, Yasmin018, Ounnunkad06, RAI2013198} including Sm-substituted hexaferrites\cite{JalliIEE06}. Ce substitution is preferred in hexaferrite as Ce is naturally abundant and is a low-cost $4f$-element. Although a few experiments have been performed with the Ce-substitution \cite{Rehman019, Chang012, PawarJMM015, MoslehJJM016}, its site preference and effect on stability and the electronic, magnetic, and optical properties remain elusive.

Experimentally, Ce-substitution has been made in the Ba-hexaferrites, {\it{e.g.}}, 20\% in Ba-site and $\sim 5\%$ in Fe-site. Although these experiments adopt a sol-gel approach, sample preparations differ. Mosleh {\it et al.}\cite{MoslehJJM016} used Fe(NO$_3$)$_3$,
Ba(NO$_3$)$_2$, CeO$_2$, ethylenediamine
tetraacetic acid (EDTA(C$_{10}$H$_{16}$N$_2$O$_8$), citric acid (C$_6$H$_8$O$_7$), and ammonia solution as precursor to make Ba$_{1−x}$Ce$_x$Fe$_{12}$O$_{19}$ $(0\leq x \leq 0.2)$, while Pawar {\it et al.}\cite{PawarJMM015} used  Ba(NO$_3$)$_2$,
Ce(NO$_3$)$_3$6H$_2$O and Fe(NO$_3$)$_3$9H$_2$O, and  C$_6$H$_8$O$_7$ in forming BaCe$_x$Fe$_{12−x}$O$_{19}$ $(0\leq x\leq 0.3)$. Chang {\it et al.} \cite{Chang012} used slightly different precursor to synthesize BaCe$_{0.05}$Fe$_{11.95}$O$_{19}$.
These reports point to two contrasting site preferences, for instance, the Sr-site in Ref.~[\onlinecite{MoslehJJM016}] and the Fe-site in Ref.~[\onlinecite{PawarJMM015, Chang012}]. 

Mosleh {\it et al.}\cite{MoslehJJM016} found no specific trend in the magnetization and coercivity by increasing Ce concentration; for instance, they found an increase of magnetization for $x=0.1$ and then a decrease for $x=0.2$. Pawar {\it et al.}\cite{PawarJMM015} found a decrease in the magnetization and Curie temperature (T$_{\text{C}}$) as well as an increase of coercivity with an increase of Ce where the site occupancy is different. Chang {\it et al.}\cite{Chang012} report that {\color{black} the tendency increases in} anisotropy field with Ce doping. In a separate experiment, Banihashemi {\it et al.} \cite{Banihashemi2019} also synthesized Sr$_{1−x}$Ce$_x$Fe$_{12}$O$_{19}$ (x$=$0.0, 0.05, 0.10) utilizing hydrothermal process and confirmed the formation of the nano-crystalline phase of the M-type hexaferrite from x-ray diffraction (XRD) and Fourier transform infrared spectroscopy (FTIR). They identified the band gap dependency (1.37–1.51 eV) on the Ce content. However, to the best of our knowledge, no theoretical studies are available in the literature on Ce-substituted hexaferrite.

In this paper, we predict the stability and electronic, magnetic, and optical properties of Ce-substituted strontium hexaferrites by performing {\it ab initio} calculations with density functional theory (DFT). We find the following key results. i) Both full and half Ce-substituted compositions have negative formation energies and are mechanically stable. ii) Ce donates an electron to Fe($2a$), leading to significant changes in the electronic structure resulting in a metal for $x=0.5$ and a semiconductor for $x=1$. iii) The net magnetic moment and $K_1$ are improved with Ce, especially $K_1$ is enhanced 
by one order of magnitude. iv) The computed optical spectra for $x=1$ show significant anisotropy between light polarization in parallel and perpendicular directions.\\



\section{Methodology and Crystal Structure}
We used Vienna Simulation Package (VASP)\cite{vasp1,vasp2} to solve Kohn-Sham eigenvalue equations within the projector augmented wave (PAW) method
and Perdew-Burke-Ernzerhof (PBE) generalized gradient approximation (GGA) exchange-correlation functional 
including the on-site electron-correlation\cite{KressePRB99,KressePRB00}. 
Fully self-consistent charge density was obtained using plane wave total energy cut-off 500 eV for all elements and $7\times7\times1$ {\bf k}-mesh for the Brillouin zone integration. For GGA + $U$ calculations, including structural relaxation, we used Dudarev's\cite{DudarevPRB98} $U_{eff}=U-J= 3$ and $4.5$ eV for Ce($4f)$ and Fe($3d$) orbitals following the previous theoretical reports\cite{LochtPRB016,BhandariPRM021}, and Heyd–Scuseria–Ernzerhof (HSE06) calculations. We employed HSE06 with Hartree-Fock (HF) mixing parameter, $\alpha=0.25$ and the
screening parameter, $w=0.2$ \AA$^{-1}$ in the electronic and optical properties calculations\cite{HSEJCP03, HSEJCP06}. For the magnetic anisotropy constant, we used fully self-consistent noncollinear GGA + $U$ + SOC calculations {\color{black} in standard 64 atom hexaferrite unit cell. The unit cell consists of Wyckoff's positions: Sr at $2d$, Fe at $2a$, $2b$, $12k$, $4f_1$, and $4f_2$, and O at $4f$, $4e$, $6h$, and two $12k$ \cite{BhandariPRM021}}.
The full Ce-substitution at Sr $2d$-site retains the original crystal symmetry to $P6_3/mmc$, space group no. 194, while a half-Ce slightly lowers the crystal symmetry $P6m2$, space group no. 187 (hexagonal), {\color{black} including the atomic site-symmetries}. The chosen {\bf k}-mesh produces well converged MCA in M-type hexaferrite structures\cite{ChlanPRB015,BhandariPRM021}

The optimized structural parameters are given in Table \ref{tab1}.
The GGA + $U$ overestimates volume (Table \ref{tab1}) for experimentally known ${\rm SrFe}_{12}{\rm O}_{19}$\cite{ThompsonJAP94}, and a similar trend is seen in the full and partial Ce-substitutions. The Ce-substitution leads to a reduction in volume as the atomic size of Ce$^{3+}$ (1.01 \AA) is slightly smaller than Sr$^{2+}$ (1.12 \AA). Specifically, the reduction is due to a slight decrease in the out-of-plane lattice constant ($c$) (Table \ref{tab1}).\\

\begin{table}[h!]
\caption{DFT optimized lattice constants in \AA~ and volume per unit cell, V, in \AA$^3$ of Sr$_{1-x}$Ce$_x$Fe$_{12}$O$_{19}$ with GGA + $U$ calculations.}\label{tab1}
\begin{ruledtabular}
\begin{tabular}{l l l l l l}
$x$ & a & b & c & c/a & V\\
\hline
$0$  & 5.927 & 5.927 & 23.17   & 3.909 &704.91 \\
$0.5$ & 5.933 & 5.933 & 23.016 & 3.879 &701.69 \\
$1$   & 5.927 & 5.927 & 22.919 & 3.867 &697.45 \\
\end{tabular}
\end{ruledtabular}
\end{table}

\section{Formation energy and Mechanical Properties}
\subsubsection{Formation energy}
Now we discuss the formation energy of ${\rm CeFe}_{12}{\rm O}_{19}$ ($x=1$) to examine chemical stability. The magnetic order of Fe sublattices remains the same as the original Sr-hexaferrite. This is confirmed by total energy calculations of all possible spin configurations in all Ce-doped compositions. All the calculations hereafter correspond to the Sr-hexaferrite type magnetic state unless explicitly stated.
It is well-known that CeO and Fe$_2$O$_3$ both exist in crystalline form\cite{LEGER1981261,ShiraneJPSJ62}. 
We compute the formation energy of ${\rm CeFe}_{12}{\rm O}_{19}$, using the relation
\begin{equation}
E_f({{\rm CeFe}_{12}{\rm O}_{19}}) = E({{\rm CeFe}_{12}{\rm O}_{19}}) - E({\text{CeO}}) - 6E({\text{Fe}}_2{\text{O}}_3), \label{eq1}
\end{equation}
where $E$(${\rm CeFe}_{12}{\rm O}_{19}$), $E$(CeO), and $E$(Fe$_2$O$_3$) are total energies of ${\rm CeFe}_{12}{\rm O}_{19}$, CeO, and Fe$_2$O$_3$ per formula unit (f.u.), which yields $E_f= -3.31$ eV per f.u. Alternatively, we compute the formation energy with respect to Ce$_2$O$_3$, Fe$_2$O$_3$, and FeO using the expression
\begin{eqnarray}
E_f(\text{CeFe}_{12}{\rm O}_{19})& = &\dfrac{1}{2}[2E({\rm CeFe}_{12}{\rm O}_{19}) - E(\text{Ce}_2\text{O}_3) \nonumber \\
& &- 11E(\text{Fe}_2\text{O}_3)- 2E(\text{FeO})],
\end{eqnarray}
and find -2.08 eV per f.u. In both cases, ${\rm CeFe}_{12}{\rm O}_{19}$ is chemically stable. 
Further, we compute $E_f$ relative to CeO$_2$ phase as
\begin{eqnarray}
E_f &=& 2E(\text{Ce}\text{Fe}_{12}\text{O}_{19}) - E(\text{CeO}_2) \nonumber \\ & & - 12E(\text{Fe}_2\text{O}_3) - E(\text{Ce}),
\end{eqnarray}
%
which is -3.88 eV /f.u.\\
We also examine the formation energy for $x=0.5$ with respect to SrO, CeO, and Fe$_2$O$_3$ as
\begin{eqnarray}
E_f(\text{Sr}_{0.5}\text{Ce}_{0.5}\text{Fe}_{12}\text {O}_{19}) & = & E(\text{Sr}_{0.5}\text{Ce}_{0.5} \text{Fe}_{12}\text {O}_{19})\nonumber \\ 
& & \hspace{-2cm} -0.5 E(\text{CeO}) - 0.5 E(\text{SrO}) - 6 E(\text{Fe}_2\text{O}_3)
\end{eqnarray}

and find $E_f=-2.94 $ eV /f.u. It is higher than for $x=1$ obtained with a similar expression, meaning a half Ce-substituted composition is chemically less stable.

To clarify Ce site-preference in Fe-sites, we compute the self-consistent total energy for {\color{black} (Sr$_2$Ce$_1$Fe$_{23}$O$_{38}$, $x \sim 0.04$)} one Ce substituted for one Fe atom in $4f_2$, $4f_1$, $12k$, $2b$, and $2a$ sites in GGA + $U$, in fully relaxed geometries. The computed energies are 0, 0.51, 1.03, 1.06, and 1.66 in eV relative to the $4f_2$ site, which indicates $4f_2$ as the most favorable Fe site. Interestingly Ce enhances the unit cell volume by $\sim 2\%$ as compared to Sr-hexaferrite, presumably due to its larger ionic radius. In contrast, the volume shrinks by $\sim 1\%$ if Ce replaces Sr. As noted earlier, the Sr has a larger ionic size, in which Ce fits relatively easily, with minimal distortion of crystal structure, while in Fe-sites, Ce creates more spacing that leads to lattice expansion.

To further scrutinize the site preference,  we compute the formation energy of one Ce substituted Sr-hexaferrite in the lowest energy Fe site, i.e., $4f_2$, by using the relation
\begin{eqnarray}
 E_f(\text{Sr}_2\text{Ce}\text{Fe}_{23}{\text O}_{38}) & = & E(\text{Sr}_2\text{Ce}\text{Fe}_{23}{\text O}_{38}) -E(\text{Ce}\text{O}_2) \nonumber \\ 
 & &\hspace{-2cm} -2E(\text{SrO}) - 11E(\text{Fe}_2\text{O}_3) -E(\text{FeO}).
 \end{eqnarray}
It gives -1.88 eV/f.u., significantly larger than Ce at Sr-site. The formation energy of other compositions for Ce at other remaining Fe-sites is even larger as the total energy is larger than for Ce at the 4$f_2$ site. The formation energy results, therefore, confirm that Ce has a strong Sr-site preference over the Fe-sites. 

 Experimentally Ce is doped in Ba-hexaferrites where the site preference is much debated. The experimental reports show that site occupancy depends on how the sample is synthesized. Indeed, there are some differences in sample preparation. For example, Mosleh {\it et al.}\cite{MoslehJJM016} used slightly different experimental precursor
than Pawar {\it et al.}\cite{PawarJMM015}. 
Both experiments report lattice parameters obtained with x-ray diffraction (XRD), which behave oppositely to the pristine system. Ce at the Ba site is shown to reduce the cell volume in Ref. ~[\onlinecite{MoslehJJM016}]. In contrast, in Refs. [\onlinecite{PawarJMM015}] and \onlinecite{Chang012}], the authors show an increase in volume with Ce at the Fe site. These trends are consistent with our results.

 


\subsubsection{Mechanical and dynamical stability}
\begin{figure}
\includegraphics[width=6cm,height=6cm]{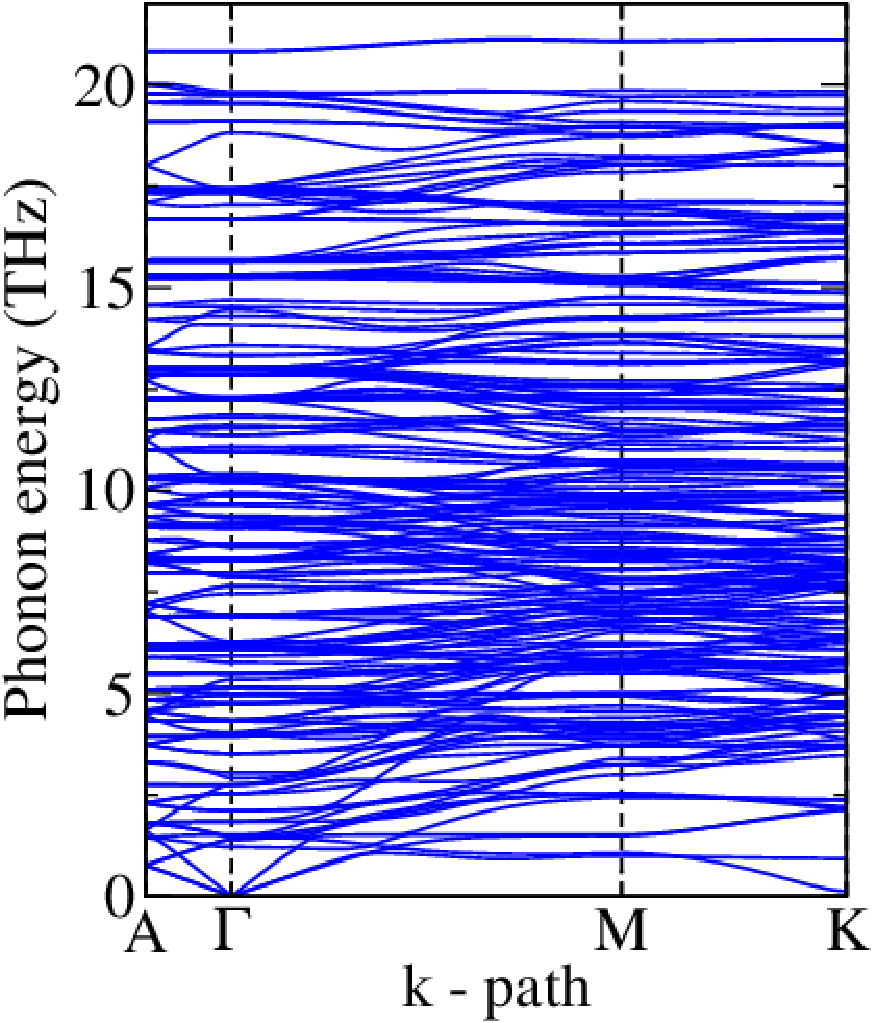}
   \caption{Phonon dispersion in Sr-hexaferrites computed along the high symmetry {\bf k}-points with GGA + $U$.\label{phonon}}
\end{figure}

Mechanical stability is described by calculating the stiffness tensor using the energy-strain relationship, which, in Voigt
notation, is
\begin{equation}
    E= \dfrac{V_0}{2}\sum_{i=1}^6\sum_{j=1}^6C_{ij}\epsilon_i\epsilon_j,
\end{equation}
 where $V_0$,  $C_{ij}$, and $\epsilon_i$ are the volume of unit cell, the elastic constants, and the components of strain tensor, respectively. As the crystal structure of Ce-substituted hexaferrites is hexagonal (space group $P6_3/mmc$), its stability can be described by 
five independent non-zero elastic constants $C_{11},  C_{12},  C_{13}, C_{33}$, and $C_{44}$.
\begin{table}
\centering
\caption{Elastic constants ($C_{ij}$), bulk modulus (K), and shear modulus (G) in Voigt notation (in units of GPa) of Sr$_{1-x}$Ce$_x$Fe$_{12}$O$_{19}$ obtained with GGA $+U$ calculations.\label{mechanical}}
\begin{ruledtabular}
\begin{tabular}{l|l|l | l}
  $C_{ij}$  & $x=1$ & $x=0.5$ & $x=0$\\
\hline
  $C_{11}$    &  297.932  &    226.311  & 283.214 \\
  $C_{12}$    &  130.157  &    109.500  & 143.012 \\
  $C_{13}$    &  110.131  &    84.067   & 102.011 \\
  $C_{33}$    &  267.509  &    203.541  & 248.651 \\
  $C_{44}$    &  71.192   &    43.796   & 65.769  \\
  K           &  173.801  & 134.603     & 167.461 \\
  G           &  79.422   &54.435       & 71.698  \\
\end{tabular}
\end{ruledtabular}
\end{table}
The elastic stiffness tensor (not given here) is the definitive positive and has all six eigenvalues positive for all compositions.
All three elastic stability criteria for hexagonal crystal \cite{MouhatPRB014}, {\it{e.g.}}, 
 $C_{11} > |C_{12}|$,
$2C_{13}^2 < C_{33}(C_{11} + C_{12})$, and 
$C_{44} > 0$ are satisfied by the computed elastic constants (Table \ref{mechanical}) for substituted as well as pure compositions confirming their mechanical stability. Interestingly, bulk modulus (K) and shear modulus (G) are similar for $x=0$ and $x=1$ compositions while smaller for $x=0.5$. These results suggest that a half-Ce-substitution makes it less stiff, likely due to the peak of the density of states at the Fermi level, which we discuss later. The zone-center ($\Gamma$-point) phonon frequencies are all positive, indicating that these are likely dynamically stable. Currently, we succeed in the full Brillouin zone phonon dispersion calculations, as shown in Fig. \ref{phonon}, for pure strontium hexaferrite, and we find no soft-mode frequencies that confirm the dynamical stability. We believe Ce and La substituted-hexaferrite, {\color{black} where an extra electron is added to the system}, presumably exhibit similar phonon behaviors at finite {\bf k}-points. However, a thorough study of phonon behaviors {\color{black} and thermodynamic stability} with element substitutions in hexaferrites requires computationally time-consuming calculations, which is worth investigating in the future.\\

\begin{figure}
\includegraphics[scale=0.24]{bandCeM.eps} \hspace{0.65cm}\includegraphics[scale=0.24]{testhalf.eps}
\\ \vspace{0.3cm} \includegraphics[scale=0.255]{sitedos.eps}\hspace{0.25cm}\includegraphics[scale=0.195]{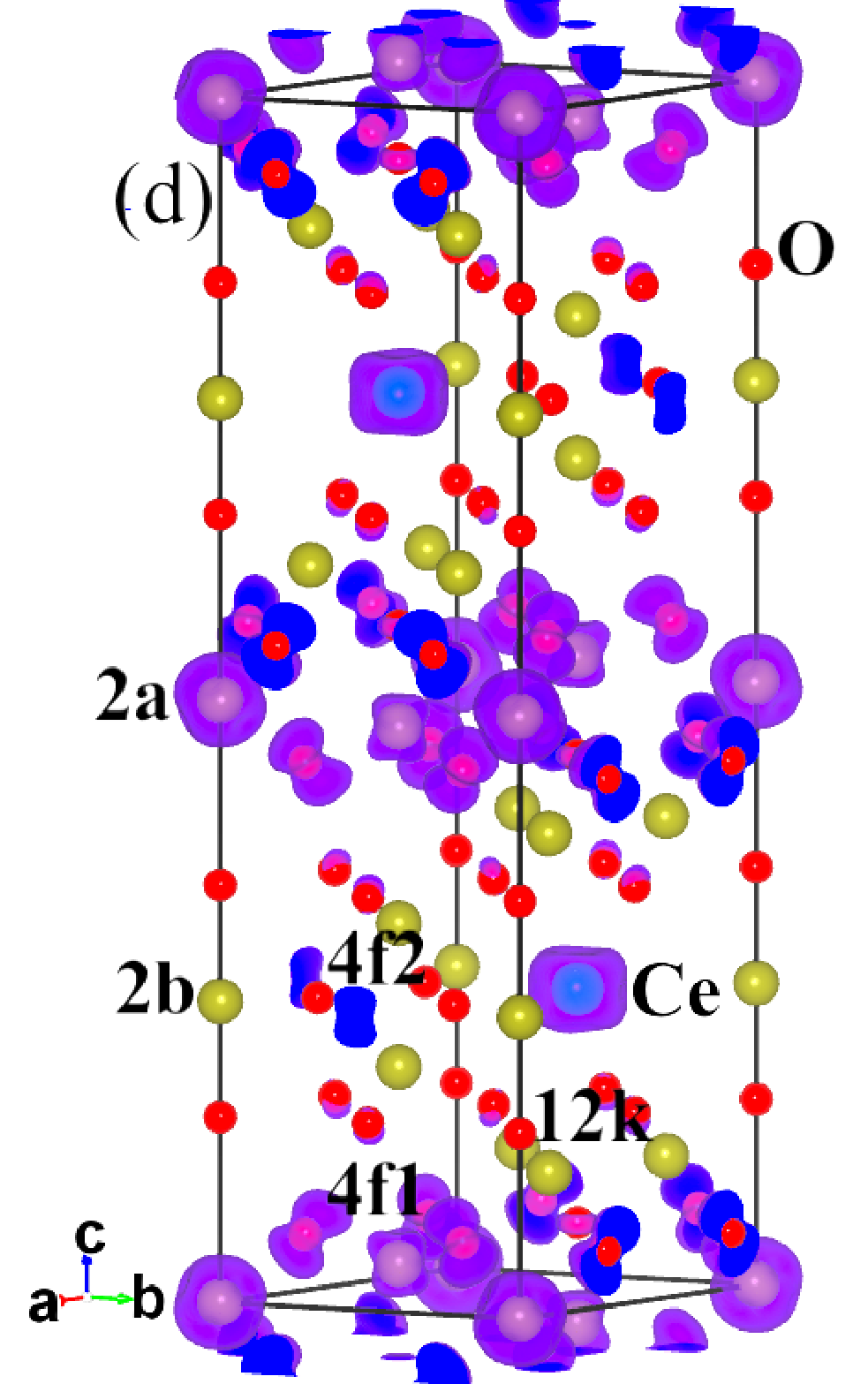}
\caption{GGA + $U$ + SOC computed band structure of Sr$_{1-x}$Ce$_x$Fe$_{12}$O$_{19}$ for (a) $x=1$ and (b) $x=0.5$. (c) Comparison of partial density of states (PDOS) of individual atoms for $x=0.5$ ({\it in black}) and $x=1$ ({\it in red}) {\color{black} obtained with GGA + $U$.} (d) Electron charge density contour occupying the valence bands near the Fermi level (in e/\AA$^3$) for $x=1$. The occupied region of PDOS corresponds to the net charge (1 e) occupying a single $4f$-orbital of Ce, confirming the formation of Ce$^{3+}$. The Fe($2a$)-PDOS occupied by transferred electrons is shown in {\it bottom panel} (c). Consistent with the electronic structure of $x=1$, the charge contours indicate the localization of Ce-substituted extra electrons near the Fe($2a$) and the localization of one electron left in Ce($4f$)-orbitals. The shape of the charge contour of electrons at the Ce atom is polarized along the $c$-axis, which is responsible for enhancing MCA. For $x=0.5$, the same is true except Fe($2a$)-$3d_z^2$ is a half-filled that leads to a metal (b). The $3d_z^2$ bands split into two manifolds as the net charge transfer slightly differs between two Fe($2a$) due to the proximity effect.}
\label{fig:dos}
\end{figure}

\section{Electronic structure}

The band structure for $x=1.0$ is shown in Fig. \ref{fig:dos}(a). The GGA + $U$ + SOC predicts an indirect band gap of 0.5 eV analogous to La-substitution\cite{BhandariPRM021}. The reduction in the band gap is due to the appearance of a localized band in the gap region of the parent compound.
The narrow band corresponds to the $3d_z^2$ orbital of Fe($2a$) as shown in Fig. \ref{fig:dos} (c) ({\it in red}). Besides, there are extra new features in band structure with varying Ce. 
First, four occupied $ 3d$ bands show up slightly above the highly localized Ce($4f$) band. Interestingly, these $3d$ bands are pushed down to the $4f$ band for $x=0.5$. 
\begin{figure}
\includegraphics[scale=0.25]{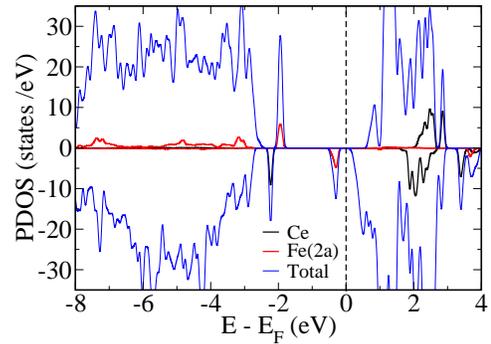}
\caption{Partial density of states (PDOS) calculated using HSE06 for $x=1$ showing a band gap and electron localization.}
\label{fig:dosHSE}
\end{figure} 
Second, the band structure [Fig. \ref{fig:dos}(b)] shows a metallic state for $x=0.5$ as the Fermi level lies in the middle of the spin-polarized (spin-$\downarrow$) band and PDOS (c) {\it in black}. 
The half-filled bands are split into two, likely due to a disproportionate amount of charge transfer from Ce, as the two Fe($2a$) are not at the equal distance from Ce. Interestingly, the half-filled bands are more dispersive along the planar distance, {\color{black} indicating asymmetry in the electron's effective mass, which may lead to interesting transport properties.} 

To better understand electron-electron correlation in $3d$ and $4f$ states, we perform more advanced HSE06 calculations for $x=1$. We show HSE06 calculated total and partial DOS of Ce and Fe($2a$) in Fig. \ref{fig:dosHSE}. For completeness, we compute the band gap of the pure system: HSE06 yields 2.4 eV, which is slightly higher than the experimental (1.9 eV) and GGA + $U$ (1.74 eV) values. We note that $U$ for Fe($3d$) is well established in the hexaferrites\cite{ChlanPRB015, BhandariPRM021}, while for Ce$(4f)$, it is unknown priori. With the help of HSE06 results, we repeat the GGA + $U$ calculations by fine-tuning $U_{eff}$ values for both $4f$ and $3d$ orbitals. The GGA + $U$ results correspond to the optimal $U_{eff}$ value, as stated in the method section (II). Qualitatively, both the GGA + $U$ and the HSE06 yield a similar band gap. {\color{black} Not surprisingly, the theoretical indirect band gap is smaller than the experiment. The experimental band gap of (1.31-1.51 eV) is a direct optical band gap, as it was obtained from a linear fit of $\alpha^2(h\nu)$ vs. $h\nu$. The calculated GGA + $U$ + SOC direct band gap of $~\sim 0.8$ eV at $\Gamma$ is also underestimated from the experiment, as discussed in Ref.~[\onlinecite{Banihashemi2019}], probably due to the presence of impurities in the low Ce-doped sample.}

To get a more quantitative picture of the localization of transferred electrons, we compute the net electrons in individual atoms by integrating the occupied PDOS in GGA + $U$. First, we analyze the valence state of Ce, which tends to fluctuate between Ce$^{3+}$ and Ce$^{4+}$\cite{ErikssonPRL88, IsnardJPA97, ShokoPRB09}.
Ce-atom consists of four valence electrons with two $6s$, one $5d$, and one $4f$, meaning it can donate two more electrons than Sr: $5s^2$. 
We find the net charge $\sim 1$e residing in Ce confirming Ce$^{3+}$ which is in agreement with Mosleh {\sl et al.'s} [\onlinecite{MoslehJJM016}] report with Fourier-transform infrared spectroscopy (FTIR) data in Ce-substituted hexaferrites.
Next, we analyze the net charge transfer to the remaining atoms. The extra electron occupies symmetry allowed\cite{BhandariPRM021} Fe at $2a$-site as given in Table \ref{tab3}. For $x=0.5$, the Fe($2a$)-atoms share an electron leading to a half-filled $3d_z^2$-orbital. On the other hand, the amount of net electron transfer in the other Fe-sites is negligible. The sharp localization of electron at $2a$-site is also confirmed by a charge density contour  [Fig.\ref{fig:dos}(c)] computed for $x=1$ within the  1 eV energy below the Fermi level.\\

\begin{table}[!htbp]
\caption{Spin and orbital magnetic moment {\color{black}(in $\mu_B$)} of individual atom in Sr$_{1-x}$Ce$_x$Fe$_{12}$O$_{19}$ ($x=0.5, 1$), the total magnetic moment per unit cell including the interstitial contribution, and the net amount of electron transfer ($q_t$) {\color{black}(in e)} to different atomic sites in units of electrons. For oxygen, only the average value is given.}\label{tab3}
\begin{ruledtabular}
\begin{tabular}{l l l l  l l l}
&$x=1$  &  & &$x=0.5$ & &\\
\hline
Atom& ~$\mu_s$ &$\mu_l$& q$_t$ &~$\mu_s$ &$\mu_l$ & q$_t$\\
\hline 
Fe($2a$)    &~3.64    &~0.01 & 1.679  &~4.02   & ~0.05 & 0.800  \\
Fe($2b$)    &~4.14    &~0.01 & 0.000  &~4.12   & ~0.03 & 0.000  \\
Fe($12k$)   &~4.24    &~0.02 & 0.031  &~4.25   & ~0.02& 0.020   \\
Fe($4f_1$)   &-4.19    &-0.02 &0.004   &-4.19  &-0.02 & 0.002  \\
Fe($4f_2$)   &-4.11    &-0.02 &0.059   &-4.13  &-0.02& 0.020   \\
O            &~0.10 & ~0.0 & 0.227 & ~0.11 & ~0.00 & 0.100\\
Ce($2d$)    &-0.27    &~2.17 &   &-0.92   & ~0.85 & ~-  \\
Total     &~37.41   &~4.48  & 2.0  &~38.01  & ~1.09 & 1.0  \\
\end{tabular}
\end{ruledtabular}
\end{table}

\section{Magnetic properties}
Now, we discuss the magnetic moments as presented in Table \ref{tab3}.
The spin moments of all Fe-atoms are, to some extent, the same as in ${\rm SrFe}_{12}{\rm O}_{19}$ except for Fe($2a$) in which the moment is reduced by about 0.5 $\mu_B$ 
for Fe$^{2+}$ ($S=2$). As the $4f$-states are less than half-filled, the signs of orbital and spin magnetic moments follow Hund's rule. For $x=0.5$,  the Ce orbital moment is significantly quenched, likely due to a stronger effect of the crystalline electric field in the metallic state. Interestingly, the spin and orbital moments nearly cancel, leading to no gain in the net magnetic moment. Indeed, experimentally, the magnetic moment is also shown to decrease for $x=0.2$\cite{MoslehJJM016}, which agrees with our results for $x=0.5$.
On the other hand, for $x=1$, Ce orbital moment (2.17 $\mu_B$) is much larger than the spin moment (-0.27 $\mu_B$), which leads to an increase of the net magnetic moment about 1 $\mu_B$ relative to ${\rm SrFe}_{12}{\rm O}_{19}$ (20 $\mu_B$).
These results suggest that Ce controls the net magnetic moment of hexaferrites.

Next, we compute T$_{\text {C}}$ by taking the energy difference between the ferromagnetic and ferrimagnetic structure of the nearest neighbor Fe-sublattices, as T$_{\text {C}} =  S^2\bar{J}_{ij}/k_B$ where $S=5/2$ is spin of each Fe and $\bar{J}_{ij}$ is 
average nearest neighbor exchange interaction. The calculated T$_{\text{C}}$ are 825.53 K and 777.74 K for $x=1$ and 0.5, which are slightly smaller than for ${\rm SrFe}_{12}{\rm O}_{19}$ (893 K). The predicted T$_\text{C}$ for CeFe$_{12}$O$_{19}$ (CeM) is slightly smaller than for ${\rm LaFe}_{12}{\rm O}_{19}$ (851 K), possibly due to the neglect of the exchange interaction between Ce$^{3+}$ and Fe$^{3+}$ in {\color{black} our oversimplified} mean-field approximation calculations.

We further calculated the T$_\text{C}$ using the static linear-response approach for CeM and LaFe$_{12}$O$_{19}$(LaM), as implemented in a Green's function  (GF), the linearized muffin-tin orbital (LMTO) method\cite{SchilfgaardeJAP1999,Questaal020} in conjunction with the atomic sphere
approximation (ASA) and the local density approximation (LDA)  and $U$ ($U_{eff}=3$ eV for Ce and 4.5 eV for Fe). The effective exchange interaction $J_0(i)$ for magnetic ion at site $i$ is obtained by using the relation $J_0(i) =\sum_{j}J_{ij}$, where $j$ is the number of the nearest neighbor atoms and $J_{ij}$ is the pairwise exchange coupling for Heisenberg Hamiltonian $H=-\sum_{i,j}J_{ij}\hat{\bf e_i}.\hat{\bf e_j}$, $\hat{\bf e_i}$ and $\hat{\bf e_j}$ are unit vectors along the direction of local magnetic moments. Then, T$_\text{C}$ is computed using the mean-field approximation, {\it i.e.}, T$_C=\dfrac{2}{3}\sum_i J_0(i)/k_B$. The computed nearest neighbor $J_{ij}$ are given in Table \ref{exchange}, which correctly show the ordering of magnetic ions, i.e., positive and negative values for antiferromagnetically and ferromagnetically aligned spins. As expected, the leading magnetic exchange-couplings are found between the nearest neighbor Fe$^{3+}$ ions at $2a-4f_1$, $2a-12k$, $2b-4f_2$, $2b-12k$, $12k-4f_1$, $12k-4f_2$, and $12k-12k$ sites, while the exchange couplings with Ce$^{3+}$ are negligible.


\begin{table}[]
    \centering
    \caption{Exchange coupling (in meV) between the nearest neighbor magnetic ions in Ce-hexaferrites computed from GF-ASA in the LDA + $U$. Fe atoms in different sites are labeled by corresponding Wyckoff positions.\label{exchange}}
    \begin{ruledtabular}
    \begin{tabular}{c|c|c |c |c |c |c}
&Ce& $2a$ & $2b$ & $12k$ &$4f_1$ & $4f_2$\\
\hline 
Ce  &  0.00  & 0.01 & 0.97 & 0.36 & 0.29 & 0.00\\
$2a$  &  - & -0.12 & 0.00 & -3.50 & 19.43&0.00 \\
$2b$ &  - &  - & -0.98 & -9.36 & 0.80 & 24.08\\
$12k$ & - & - & - & -10.34 & 18.23 & 15.24 \\
$4f_1$ & - & - & -& - & -2.91 & -0.57 \\
$4f_2$& - & - & - & - & - & -0.52\\
\end{tabular}
  \end{ruledtabular}
\end{table}
We found T$_{\text{C}} = 905$ K for LaM and 914 K for CeM, which agree qualitatively with respective values obtained with an oversimplified approach. For CeM, T$_\text{C}$ obtained by including the exchange interactions of Ce$^{3+}$ with Fe$^{3+}$ is slightly larger. Overall there are no significant differences in T$_\text{C}$ with La and Ce substitutions. \\

\section{Magnetocrystalline anisotropy}
\begin{table}
\caption{GGA + $U$ + SOC calculated magnetic anisotropy constant, $K_1$, in Sr$_{1-x}$Ce$_x$Fe$_{12}$O$_{19}$ ($x$ is a variable $0$ to $1$) and its comparison with experiment and previous theory for Sr/LaFe$_{12}$O$_{19}$. The values are presented in units of MJ/m$^3$ for $U_{eff}=2-4$ for Ce and $4.5$ eV for Fe.}\label{tab4}
\begin{ruledtabular}
\begin{tabular}{l l l l l}
$K_1$ & $x=0$ & $x=1$ & $x=0.5$ & ${\rm LaFe}_{12}{\rm O}_{19}$\\
\hline
This work\footnote{$U_{eff}=4$ eV for Ce} &0.35 & 10.31 & 2.70 & 0.66\\
This work\footnote{$U_{eff}=3$ eV for Ce} &0.35 & 8.14 & 1.71 & 0.66\\
This work\footnote{$U_{eff}=2$ eV for Ce} &0.35 & 6.34 & 1.14 & 0.66\\
Expt.\cite{JahnPSS69,ShirkJAP69} & $0.35-0.36$ &- &- &$0.5-0.8$  \\
\end{tabular}
\end{ruledtabular}
\end{table}

Now we turn our discussion to magnetic anisotropy. To predict MCA, we employ a fully self-consistent relativistic non-collinear scheme to compute the total energy in two different crystalline directions, viz., $E_{001}$ (magnetocrystalline easy) and $E_{100}$ (magnetocrystalline hard) along the crystalline $c,~i.e., [0,0,1]$  and $a,~i.e, [1,0,0]$ directions, including  SOC in GGA + $U$. The magnetic anisotropy energy ($MAE$) is then obtained as $MAE=E_{100}-E_{001}$ from which we evaluate $K_1$ as $K_1=MAE/V$, where $V$ is the unit cell volume.

\begin{figure}
    \centering
    \includegraphics[scale=0.275]{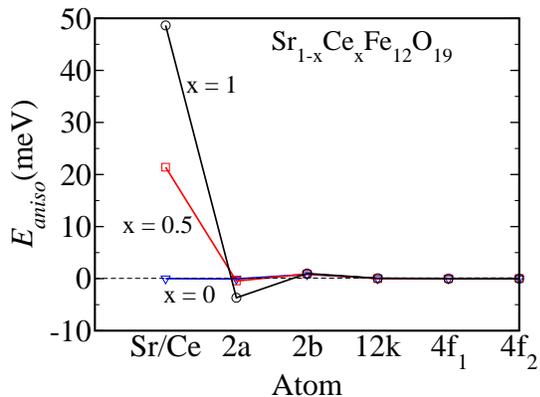}\\
    \caption{On-site anisotropy energy ($E_{aniso}$) contributions. The Ce has the highest $E_{aniso}$ contribution. Fe atoms are labeled by their corresponding Wyckoff position. The onsite Fe contribution is positive at $12k$ and $2b$ and negative at $2a$. The contribution is negligible at $4f_1$ and $4f_2$-sites.}
    \label{SOCanisotropy}
\end{figure}

{\color{black} $K_1$ is computed for different values of $U_{eff}$ for Ce(4$f$) [Table \ref{tab4}]. For $U_{eff}\sim 3$ eV, the computed value of $K_1$ in 100\% Ce-doped hexaferrite is nearly 8 times larger than in strontium-hexaferrite.} These results clearly show a big influence of $4f$-electrons. Qualitatively, the effect of $4f$-electrons can be understood by the charge density contours of occupying states near the Fermi level. As seen from Ce's prolate type of charge density contour, it costs huge energy to rotate it in other directions. The most direct quantitative evidence for the site-contribution to MAE can be obtained from the energy difference between easy and hard axes (anisotropy energy) viz., $E_{aniso}=E_{soc}[100]-E_{soc}[001]$, where $E_{soc}[001]$  and  $E_{soc}[100]$ are atomic SOC energies for magnetization directions $[0,0,1]$ and $[1,0,0]$. The net MAE is obtained from the expression\cite{SunPRB020} $\dfrac{1}{2}\sum_{i}E_{aniso}^i$, where $i$ denotes the atomic site.
Figure \ref{SOCanisotropy} shows the $E_{aniso}$ to MAE for different Ce substitutions. For $x=0$, 
the MAE is primarily governed by Fe($2b$) and Fe($12k$) with some contribution from Fe at $4f_1$ and $4f_2$. Although small, Fe($2a$) gives a planar contribution.
Interestingly, electron doping to Fe($2a$) results in uniaxial or planar MAE depending on the substituted elements. For instance,
it produces a uniaxial contribution and is a leading term in La substituted hexaferrite.
In CeM, the shape of the $3d_z^2$ orbital is significantly distorted [Fig. 1(d)] from the $c$-direction, unlike in LaM, probably due to the $4f$ electrons. The Fe($2b$) site contribution is uniaxial and remains unchanged after substitutions. The $4f_1$ or $4f_2$ contribution changes sign; however, it does not affect the MAE since it is much smaller.
Of note, as expected in $4f$-electron systems, the Ce-site contribution is larger by orders of magnitude than other sites.

A similar $4f$-electrons effect in MAE is found for $x=0.5$, while the increase is not proportional to Ce. In the metallic system, the nonlinear trend in MAE is expected;  because the itinerant electrons can perturb the $4f$ charge distribution and affect the crystalline electric field. Indeed, the Ce-site contribution to MAE [Fig. \ref{SOCanisotropy}] is reduced, consistent with the reduced Ce-orbital magnetic moment. Lastly, the Fe-sites contribution to MAE is less significant.\\

\begin{table}
\caption{Calculated magnetic properties and comparison with available experiments {\color{black}(for $x=0$) at room temperature otherwise stated} for Ce-substituted compositions (Sr$_{1-x}$Ce$_x$Fe$_{12}$O$_{19}$)  with {\color{black} $U_{eff}=3$ eV} for Ce and $4.5$ eV for Fe viz., magnetic anisotropy constant ($K_1$), saturation magnetization ($M_S$), anisotropy field ($H_a$), BH$_{max}$, and magnetic hardness parameter ($\kappa$) (units are given inside the parenthesis).}\label{tab5}
\begin{ruledtabular}
\begin{tabular}{l l l l l}
 parameter & $x=0$ & $x=1$ & $x=0.5$ & Expt\cite{CoeyIEEE011}\\
\hline
$K_1$ (MJ/m$^3$)  &0.35 & 8.14 & 1.71 & 0.35 \\
$M_S$ (T)      & 0.60 & 0.64 & 0.59 &0.31\footnote{Experiment at 1100 $^{\circ}$ C for $x=0.1$ Ref.\onlinecite{MoslehJJM016}}, 0.48 - 0.59\footnote{ Experiment at 5 K  for $x=0$ Ref.~[\onlinecite{SeifertJMMM09}]} \\
$H_a$ (KOe)  &14.58 & 320.04 & 72.56 & $14.88^a$-20.01\\
BH$_{max}$ (kJ/m$^3$) & 72.42 & 81.12 & 69.82 & 45-$69.53^a$\\
$\kappa$ & $1.13$ & 5.01  &2.47 & 1.35 \\
\end{tabular}
\end{ruledtabular}
\end{table}

Table \ref{tab5} summarizes the computed magnetic properties of Ce-substituted compositions and compares them with experimental values of pure Sr-hexaferrite. The saturation magnetization ($M_S$) is enhanced in full Ce-hexaferrite. However, 50\% Ce substitution does not increase the $M_S$. {\color{black} Our calculated $M_S$ lies within the range of  experimental values, which vary from 0.31  to 0.59 T depending on the temperature, sample preparation techniques, and micromagnetic structures.} 
Magnetic anisotropy field ($H_a$) can be estimated using the relation 
\begin{equation}
 H_a = \dfrac{2K_1}{M_S}.   
\end{equation}
For Ce-substituted compositions, $H_a$ enhances by about a factor of 10, consistent with the predicted value of $K_1$. 

BH$_{max}$ which is estimated from $M_S$ as 
\begin{equation}
\text{BH}_{max} = \dfrac{\mu_0 M_S^2}{4}.  
\end{equation} BH$_{max}$ is the largest for full Ce. Magnetic hardness parameter $\kappa$ 
is given by 
\begin{equation}
 \kappa = \sqrt{K_1/\mu_0 M_S^2},   
\end{equation}
and it satisfies the criterion\cite{CoeyIEEE011}  ($\kappa > 1$) of hard magnet for all compositions. The full Ce substitution of Sr produces the hardest permanent magnet with the largest value of $\kappa$.
\\

\section{ Optical anisotropy}

To understand the effect of Ce($4f$)-states in optical properties of CeFe$_{12}$O$_{19}$, we compute the optical absorption coefficient
\begin{figure*}[!htbp]
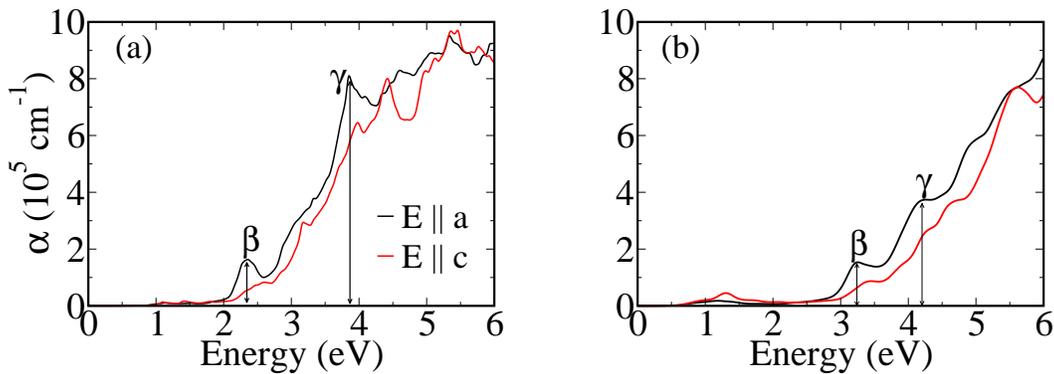

    \includegraphics[scale=0.25]{alpha.eps} \hspace{1cm}
    \includegraphics[scale=0.25]{HSEalpha.eps}
    \caption{Optical absorption coefficient, $\alpha(\omega)$, of full Ce-substituted hexaferrite ($x=1$), calculated with GGA + $U$ + SOC (a) and HSE06 + SOC (b). The strength of dipole-transition between $4f$ states is rather strong  compared to $3d$-states, resulting in two peaks $\beta$ and $\gamma$ at $\sim 2$ and $\sim 4$ eV in GGA + $U$ + SOC. These peaks shift by $\sim 1$ eV with HSE06. In the HSE06 results, an additional peak appears in the low energy region around 1.3 eV for $E\parallel c$.\label{alpha}}
\end{figure*}
$\alpha (\omega) =  \dfrac{\omega  \varepsilon_{2}(\omega)}{c n(\omega)} ,
$ where $c$, $\omega$, and  $n (\omega) = \dfrac{1}{\sqrt{2}}[\varepsilon_1(\omega) +\sqrt{\varepsilon_1(\omega)^2 +\varepsilon_2(\omega)^2}]^{\frac{1}{2}}$, and $\varepsilon_{2}(\omega)$ are speed of light, photon energy, refractive index, and the imaginary part of the macroscopic dielectric constant from which the real part of the macroscopic dielectric constant $\varepsilon_{1}(\omega)$ is obtained using Kramers-Kronig transformation. The $\varepsilon_{2}(\omega)$ is calculated from the expression

\begin{equation}
\varepsilon_{2}(\omega) =   \frac{4\pi^2e^2}{m^2\omega^2}\sum_{v, c}  \int_{BZ}  \frac {d^3 k} {( 2\pi)^3}      
| \hat e \cdot {\bf M}_{cv} ({\bf k})|^2       \delta(\hbar \omega-\epsilon_{c{{\bf k}}}+\epsilon_{v{{\bf k}}}),
\label{epsilon}
\end{equation}
where  $\hat e$ and $ {{\bf M}}_{cv} ({{\bf k}}) = \langle \psi_{c{{\bf k}}}|  {\bf p} |\psi_{v {\bf k} }\rangle$ are the
light polarization vector and  the momentum matrix element between the conduction and the valence states. 
We find significant changes in the $\alpha(\omega)$ as compared with ${\rm LaFe}_{12}{\rm O}_{19}$, where vertical transitions occur between Fe($3d$) states as discussed in our previous paper\cite{BhandariPRM021}.
Overall the intensity of $\alpha(\omega)$ is enhanced considerably with Ce, including the appearance of two main peaks labeled by $\beta$ and $\gamma$ in Fig. \ref{alpha}. It is expected as the $4f$-states are located near the Fermi-level, for which the strength of matrix elements of dipole moments corresponding to the transition from the occupied $4f$ to unoccupied $4f$-states is much stronger. The unoccupied $4f$-states are split into two manifolds because of the combined effects of crystal-field, exchange, and SOC. $\beta$-peak corresponds to the transition from the occupied $4f$ to lower unoccupied at $\sim 2$ eV, and $\gamma$ corresponds to upper unoccupied $4f$-states at $\sim 4$ eV. The optical absorption is spatially localized to Ce-atom, which could be interesting to examine experimentally. The other interesting result is enhanced anisotropy in optical absorption. Indeed, the real parts of dielectric constant $\varepsilon_1$ are 6.94 for $E\parallel a$ and 6.18 for $E\parallel c$ consistent with $\alpha(\omega)$.

HSE06 calculations reveal qualitatively similar optical spectra, except now that the absorption peaks are shifted. In particular, $\beta$-peak is shifted by about 1 eV, while $\gamma$-peak shifts by about 0.5 eV. These are consistent with the alignment of Ce($4f$)-bands relative to the Fermi level. The height of $\beta$-peak is similar, while $\gamma$-peak is reduced by about 2. 
Interestingly, HSE06 + SOC produces another peak of about 1.3 eV corresponding to the interband transition between the Fe($3d$)-states for $E\parallel c$, which is much weaker with the GGA + $U$ + SOC calculations. The magnitudes of $\varepsilon_1$ are 3.94 for $E\parallel a$ and 3.65 $E\parallel c$. The difference in the dielectric constants between the two approaches is expected, partly due to the use of different numbers of virtual states in the calculations. Unfortunately, the convergence of $\varepsilon$ in HSE06 calculations is always problematic due to the huge computational cost.
However, both approaches yield a similar trend in the anisotropy.

\section{Metastable sites and magnetism}
\begin{figure}
    \centering
    \includegraphics[scale=0.275]{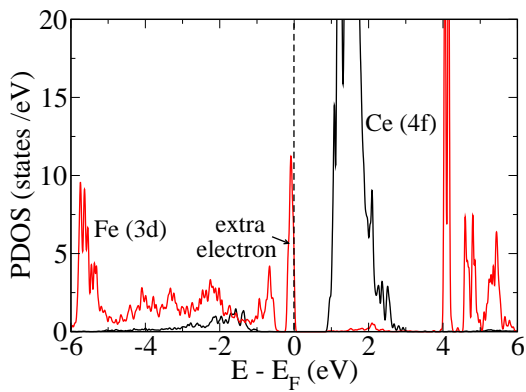}
    \caption{PDOS of Ce($4f$) and Fe($3d$) at $2a$ (the nearest to the Ce in the unit cell) computed with GGA + $U$ + SOC for Ce-substituted at Fe-$4f_2$ site. The extra Ce electron transfers to the Fe($3d$) states localized below the Fermi level.}
    \label{fig:ceat4f2}
\end{figure}

 By performing the GGA + $U$ + SOC calculations, we examine the effect of Fe-site substitutions by Ce in magnetic properties and local spin quantization. Depending on the Ce-site occupancy, the net magnetic moment changes from 34 $\mu_B$ to 44 $\mu_B$, as given in Table \ref{tab6}. To illustrate it, we discuss the PDOS of Ce and Fe (nearest to Ce) while replacing one Fe at $4f_2$ (the most favorable Fe site) in Fig. \ref{fig:ceat4f2}. The $4f$-states are delocalized below the Fermi level. Quite interestingly, one electron of Ce transfers only to the nearest Fe($2a$), forming a localized $3dz^2$ state. It reduces Fe$^{3+}$ to Fe$^{2+}$, as found in Ce at Sr-site.  
Ce remains to be Ce$^{3+}$, which helps to increase the net magnetic moment to 44 $\mu_B$ for the following reasons: (i) the negative spin magnetic moment contribution of about 5 $\mu_B$ at $4f_2$ of one Fe-atom vanishes and (ii) the magnetic moment of Ce itself is much smaller $\sim$ 1 $\mu_B$. A similar situation holds in $4f_1$-site. At the $2b$ site, Ce gives an extra electron to Fe($2a$), producing Fe$^{2.5+}$ and further reducing the net moment to 34 $\mu_B$. Interestingly, at the $12k$ site, our calculations show that the Ce donates an extra electron to the nearest Fe($12k$), resulting in a similar net moment.

Next, we {\color{black} calculate total energies by choosing the spin quantization directions along [0,0,1], [1,0,0], and [1,1,1] using self-consistent GGA + $U$ + SOC method and} investigate the Ce-induced spin-canting by analyzing the $E_{aniso}$ and $MAE$. {\color{black} Fortunately, we did not have to constrain the spins because they remained aligned with the initialized directions even after completing the full self-consistent calculations.}
The computed $E_{aniso}$ (not shown here) suggest that the local spin magnetic moment directions of Ce($4f_2$) and Fe($2a$) are not along the [0,0,1] (Gorter's type\cite{GorterPIEE57}), rather they are along the [1,1,1], while the directions of other remaining Fe-spin magnetic moments remain the same. As Ce($4f$) with a large $E_{aniso}$ primarily controls the magnetic anisotropy, the whole spin-system becomes uni-axial along the crystalline [1,1,1] direction with $MAE=E_{001}-E_{111}= 0.45$ meV.
This local spin-canting could arise due to the following reasons: (i) {\color{black} a complex spin structure around the defect} and (ii) a significant change in the crystalline electric field.  Experimentally Sm is already known to replace Sr\cite{JalliIEE06}. In a separate experimental work\cite{Yasmin018}, authors report that Sm causes a spin-canting at Fe-sites. Pawar {\it et al.} also discussed the possibility of spin canting if Ce is doped at Fe-sites[11]. Although our DFT calculations suggest Fe-spins canting near the impurity, it does not dramatically change the net magnetic moment (see Table VII) as reported in experiment\cite{Yasmin018}, in which $M_S$ is shown to decrease significantly by increasing the Ce concentration.

\begin{table}
\caption{GGA + $U$ calculated the net magnetic moments ($m$) (in $\mu_B$) for SrCeFe$_{23}$O$_{19}$ by replacing a Fe atom with Ce in different sites and the optimized lattice constants $a$ and $c$ (in \AA). }\label{tab6}
\begin{ruledtabular}
\begin{tabular}{l l l l l l l l}
 parameter & $2a$ & $2b$ & $12k$ & $4f_1$ & $4f_2$\\
\hline
a  &5.95 &5.99 & 5.98 & 5.96 & 5.96 \\
c  &23.48 &23.15 & 23.36 & 23.41 & 23.68\\
$m$ & 36 &34 & 34 & 44 & 44\\
\end{tabular}
\end{ruledtabular}
\end{table}

\section{Conclusion}
In summary, by performing DFT, including hybrid functional calculations and analyzing formation energies and elastic constants,
we uncover the lattice stability of Ce-substituted M-type strontium hexaferrites. Both full- and half-Ce-substituted compositions are chemically and mechanically stable. The full Ce results in fully occupied narrow $3d_z^2$-derived bands of Fe($2a$), yielding a reduced band gap.
On the other hand, the half-Ce leads to a half-filled $3d_z^2$-band producing a metallic state. These results, including the computed net amount of 4\textit{f} electron in Ce, confirm the formation of Ce$^{3+}$.
The Ce increases the MCA of the parent compound by at least an order of magnitude, which is a direct consequence of Ce(4\textit{f})-electrons. 
The net magnetic moment also increases with the full Ce as the net Ce-contribution results from the orbital moment following Hund's rule. In addition, due to Ce($4f$), the computed optical spectra for CeFe$_{12}$O$_{19}$ show significant anisotropy between light polarization in parallel and perpendicular directions with enhanced transitions between Ce($4f$) and conduction bands. If grown successfully, we believe the Ce-substituted M-type hexaferrite can be a potential candidate material for a high-performance permanent magnet.\\

\acknowledgements{This work is supported by the Critical Materials Institute, an Energy Innovation Hub funded by the U.S. Department of Energy, Office of Energy Efficiency and Renewable Energy, Advanced Manufacturing Office. The Ames National Laboratory is operated for the U.S. Department of Energy by Iowa State University of Science and Technology under Contract No. DE-AC02-07CH11358.} 

\bibliography{cehexa}
\end{document}